# Innovative Countermeasures to Defeat Cyber Attacks Against Blockchain Wallets: A Crypto Terminal Use Case

Pascal Urien

Telecom Paris
*19 Place Marguerite Perey 91120 Palaiseau, France*
Pascal.Urien@telecom-paris.fr

*Abstract*— Blockchain transactions are signed by private keys. Secure key storage and tamper-proof computers are essential requirements for deploying a trusted infrastructure. In this paper, we identify some threats against blockchain wallets and propose a set of physical and logical countermeasures to thwart them. We present the crypto terminal device, operating with a removable secure element, built on open software and hardware architectures, capable of detecting a cloned device or corrupted software. These technologies are based on tamper-resistant computing (javacard), smart card anti-cloning, smart card content attestation, application firewall, bare-metal architecture, remote attestation, dynamic Physical Unclonable Function (dPUF), and programming tokens as a root of trust.

## I. INTRODUCTION

Blockchain transactions are signed by private keys, which imply security requirements similar to those of electronic signature processes or EMV payments made using bank cards and payment terminals.

The main security requirements are secure key generation, secure key storage, and tamper resistance. Some threats of key theft are listed in [29].

A secret key is an integer value whose size is 32 bytes for a 256-bit elliptic curve. The best practice should be to use a true random number generator (TRNG). Nevertheless, many wallets use passphrases as seeds for key generation; they are based on a choice of words (e.g., 12 values chosen from 2048), which leads to $2^{132}$ possible combinations. The article [30] suggests recovering the passphrase by a brute-force dictionary attack. Another technique, called "brain wallet", calculates the private keys from the passphrases using a hash procedure; the article [31] demonstrates brute-force attacks.

Many blockchains, for example, Bitcoin or Ethereum, use the ECDSA (*Elliptic Curve Digital Signature Algorithm*) signature. This procedure generates a random number k, and a point on the elliptic curve (secp256k1) with the generator G, k.G= (x,y) from which is calculated an integer r= x mod n, n being the order of the curve. The knowledge of k, the reuse of k, or a known difference between two values of k, makes it possible to recover the private key. The paper [32] studied such a key leakage in Bitcoin. It extracted 647,110,920 signatures and found 1,068 distinct r-values appearing at least twice and used by 4,433 keys, or about 0.35% of the r-values. The most frequently observed value is k = 1/2 mod n. It should be mentioned that malicious random number generators, such as kleptograms [33][34][35], can be used to recover private keys after two signatures.

Even if the signatures are computed by a secure and trusted device, they can be performed by malicious software, whose purpose is to generate fraudulent transactions.

In EMV payment systems, point-of-sale (POS) terminals are equipped with security stickers and battery-powered electronics to detect attempted fraudulent use. The *PCI Security Standards Council* lists approved companies and vendors. Nevertheless, no standards are available today for assessing security in blockchain operations. In addition, low-cost wallets are sold by e-commerce stores and delivered by untrusted supply chains.

Secure elements, such as smart cards, are an effective technology for providing secure storage and trusted cryptographic operations, with high levels of security (EAL 6+ according to Common Criteria -CC- standards) and countermeasures to counter side-channel threats. However, as with EMV cards, it is necessary to detect malicious clones.

Secure elements require a terminal with additional functionalities: user interface (touch screen, etc.) and communication (USB, Bluetooth, etc.). Proof of integrity of the software executed by the terminal is an essential prerequisite. From an industrial property point of view, detecting counterfeit electronic cards is an important feature.

In this paper, we describe a dedicated blockchain terminal based on open hardware (i.e., Arduino platform) and software technologies. It incorporates a set of countermeasures capable of verifying the authenticity and integrity of the firmware and detecting cloned devices. Thanks to the open hardware, the terminal is realized with components supported by the Arduino integrated development environment (IDE), but it can also be integrated into a dedicated electronic board.

The first prototype [9] used a two-line 16-character LCD and a 4x4 keyboard. These elements were replaced by a touch screen in the second prototype [10], which also supports a *Bluetooth Low Energy* (BLE) interface. An original ISO7816 library has been developed to support smart cards by Arduino environments. The firmware integrates a set of original security features detailed in [16] [17] [19], whose goal is to provide a bare-metal architecture with a root of trust, a software attestation procedure, and hardware fingerprints.

This paper is organized according to the following outline. Section 2 briefly introduces the blockchain wallet state of the art and identifies some threats; it also presents an original algorithm to detect duplicated code shards. Section 3 describes the crypto terminal, an open device that integrates a set of countermeasures. Section 4 details the javacard security features: tamper-resistant computing, anti-cloning, content self-attestation. Section 5 presents the terminal security features: applicative firewall and bare-metal architecture. Section 6 introduces remote attestation using the BMAC algorithm. Section 7 describes static and dynamic PUF. Section 8 introduces programming tokens as a root of trust. Finally, Section 9 concludes this paper.

## II. . STATE OF THE ART

### A. Blockchain Wallet

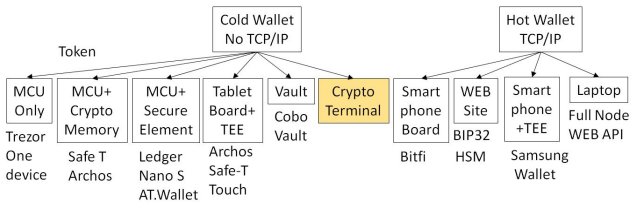

Fig1. Overview of blockchain wallets.

A blockchain wallet comprises two parts: a set of private keys used to sign transactions and a set of parameters needed to generate transactions. The secure storage and use of private keys is a major security prerequisite.

While cryptographic keys can be used without communication with the blockchain infrastructure, the information needed to build a transaction must be retrieved, which requires an Internet connection. Figure 1 illustrates two broad categories of blockchain wallets: hot wallets (with TCP/IP stack) and cold wallets (without TCP/IP stack).

The first generation of hot wallets was based on software without tamper-resistant features

```
EE 00 00 00 0B 00 01 0A 64 65 66 61 75 6C 74 6B ........defaultk
65 79 00 00 0D 00 01 0C 59 6F 75 72 20 41 64 64 ey......Your Add
72 65 73 73 28 00 01 04 6E 61 6D 65 22 31 36 69 ress(...name"16i
4C 68 37 7A 74 4C 68 32 6D 77 6B 68 53 76 5A 65 Lh7ztLh2mwkhSvZe
48 78 4B 57 51 56 56 4D 47 47 67 66 57 47 71 00 HxKWQVVMGGgfWGq.
1A 01 FD 17 01 30 82 01 13 02 01 01 04 20 38 ......0....... 8
78 AC FB 86 DD E1 B7 6D 03 1F 3A A7 F8 52 36 E6 x......m..:.R6.
EE 0F 5F 86 55 24 44 13 17 DE 31 97 14 A9 16 A0 .._.U$D...1.....
```

Fig2. Private key storage in the wallet.dat file used by the bitcoin.exe software for windows

As an example, the Bitcoin.exe software for win32 was written in 2009 by Satoshi Nakamoto [1]; it consists of about 16,000 lines of C++ code, and its binary image is 6 MB. It included information managed by a non-SQL database, the Berkeley DB. In particular, the private keys were stored in the file named wallet.dat. As shown in Figure 2, the private key (colored in yellow) is not encrypted and can be identified by its associated Bitcoin address:

"16iLh7ztLh2mwkhSvZeHxKWQVVMGGgfWGq"

In the Bitcoin system, the *coinbase* transaction is the first transaction in a block, used to transfer the potential reward to an address identified by its Hash160 attribute:

$$Hash160 = RIPEMD160(SHA256(PublicKey))$$

The bitcoin.exe software generates a random address before mining a new block. The loss or theft of the unprotected wallet.dat file was a major risk, not always understood by early miners.

Many blockchains use the *Elliptic Curve Digital Signature Algorithm* (ECDSA) signature. Given a message signature made of two integers (r,s):

$$s = k^{-1}(e + z.r) \bmod n$$

with n being the group order, e the message hash, z the private key, k a random number kG=(x,y), and r=x mod n.

And given two signatures of two different messages, $M_1$ and $M_2$, with the same r in $(r,s_1)$ and $(r,s_2)$, $e_1$=hash($M_1$), $e_2$=hash($M_2$), the private key is computed as:

$$z = (e_1.s_2 - e_2.s_1)\, r^{-1}\, (s_1 - s_2)^{-1} \bmod n$$

Therefore, ECDSA generation requires a random number generator (RNG) or a deterministic procedure based on the message fingerprint (as detailed by RFC 6979). In 2010, EC private keys were extracted from PS3 consoles by exploiting the lack of a random number generator [21]. ECDSA, therefore, relies on trusted software.

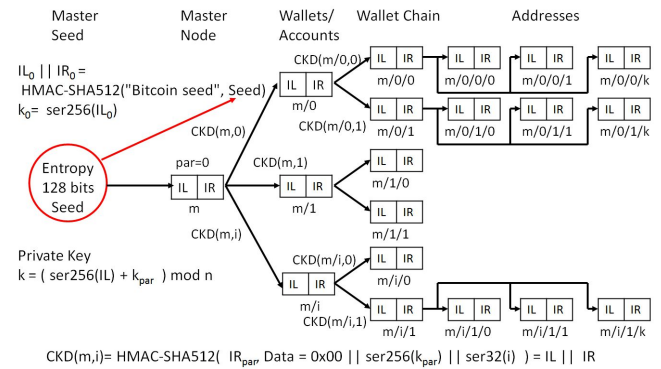

Fig3. Illustration of hierarchical deterministic wallets, according to the BIP32 specification

In Bitcoin, according to the BIP32 specification, keys are computed in deterministic wallets (see Figure 3). A root secret (512 bits), divided into two parts (IL0 and IH0), is calculated from a seed. Then, child nodes, identified by a 32-bit integer index, compute 512-bit secrets based on their parents' IL and IR attributes, using the CKD(m,i) procedure. Finally, the private keys (256 bits) are computed, from which the account addresses are derived. Therefore, an address (and the associated private key) is identified by a path in the BIP32 key tree. The suggested hierarchy is as follows: master node, wallet accounts, wallet chains and addresses.

The BIP32 seed can be computed from a passphrase according to the BIP19 specification. The passphrase is a set of words selected from a dedicated list of 2048 elements. The computation procedure is based on the *password-based key derivation function* (PBKDF2) defined by RFC 2898. This mechanism allows the recovery of deterministic wallets, as the key paths (i.e., the key identifiers) are public values. Nevertheless, if the passphrase is stolen, the private keys can be recovered by a brute-force attack on the key identifiers.

In 2014, a major Bitcoin company, Mt.Gox, went bankrupt due to a data breach in untested software, leading to the illegitimate use of private keys [29]. In addition, the privileges of system administrators allow the hacking of private keys. As an illustration, Bitfi [2] is a UNIX smartphone without a baseband chip, including an SD memory card, and offering Wi-Fi connectivity. It was rooted, and private keys were extracted.

To prevent these attacks, data centers use *Hardware Security Modules* (HSMs), whose security requirements are described in the FIPS 140-2 standard ("*Security Requirements For Cryptographic Modules*") [3]. Four security levels are defined, ranging, according to the Common Criteria (CC) terminology, from EAL1 to EAL4. Cryptographic keys are managed by tamper-resistant systems whose access requires multi-factor authentication (e.g., password, hardware key, and PIN). For example, Coinbase recommends [4] storing the PIN-protected hardware key in a vault so that its use involves two human brains, one of which knows the smart card PIN and the other the vault combination.

Some smartphone wallets are based on a hardware keystore, which runs in a TEE (*Trusted Environment Execution*) processor [5]. Most Android mobiles are equipped with such processors. TEE implements a secure computing environment; however, according to [22], the hardware techniques and processes used for smart cards do not apply to standard *System-on-Chip* (SoC) technology.

The main idea of cold wallets is the offline storage of private keys. Communication with the outside world uses USB, Bluetooth, Wi-Fi, or QR codes read by a camera and displayed on screens (e.g., cobo vault).

There are two broad categories of cold wallets: those that are fully software-based and those that include a tamper-resistant device.

The Trezor One is an all-software device with a screen, two buttons, and a USB interface. According to [6], the private keys were recovered by a single-powered attack (SPA) exploiting the unprotected implementation of the double and add algorithm, in which a public key (zG, G being a generator) is computed by parsing bit by bit, the private key z.

$$z = \sum_{i=0}^{255} b_i . 2^i, b_i \in \{0,1\}$$

An elliptic curve addition is computed for every non-null bit, which enables the bit recovery, for example, by monitoring the power consumption.

$$zG = \sum_{0}^{255} b_i . D_i, \quad D_0 = G, \quad D_{i>0} = 2.D_{i-1}$$

Some wallets use crypto memory (for example, *Safe-T Archos*) or Trusted Execution Environment (for example, *Archos Safe-T Touch*). Crypto memory devices provide secure storage; they work with mutual authentication procedures and encrypted content.

The *Ledger Nano S* stores keys in a secure element. The device comprises a microcontroller, a secure element, a screen, two buttons, and a USB interface. Upon reset, a software handler sends memory chunks to the secure element, which hashes their content, and finally checks a signature; upon success, the secure element is unlocked. This mechanism was broken [7] by using a duplicate area of memory. The reference [8] demonstrated a successful replacement of the token software.

```
610: 3f 4f    sbci  r19, 0xFF      652: 3f 4f    sbci  r19, 0xFF
612: d9 01    movw  r26, r18       654: d9 01    movw  r26, r18
614: ec 90    ld    r14, X         656: ec 90    ld    r14, X
616: 1e 25    eor   r17, r14       658: 1e 25    eor   r17, r14
xyz:                                xyz:
648: 17 96    adiw  r26, 0x07      68a: 17 96    adiw  r26, 0x07
64a: 2c 91    ld    r18, X         68c: 2c 91    ld    r18, X
64c: f2 26    eor   r15, r18       68e: f2 26    eor   r15, r18
64e: 9f 01    movw  r18, r30       690: 9f 01    movw  r18, r30
650                                 692
```

Fig4. Illustration 64 bytes duplicate code shards

As illustrated by [7], software instruction blocks may be duplicated. We call shards these blocks. We assume that shards do not include instructions that explicitly modify the program counter (PC), such as JUMP or CALL. An example of code shards is provided in Figure 4. We implemented original procedures based on *Ukkonen*'s *algorithm* for *Suffix Tree Construction* [23] to detect and remove shards. This algorithm finds the longest duplicated binary string; replicas are replaced by random numbers. Finally, it produces a list of duplicate code fragments and their size.

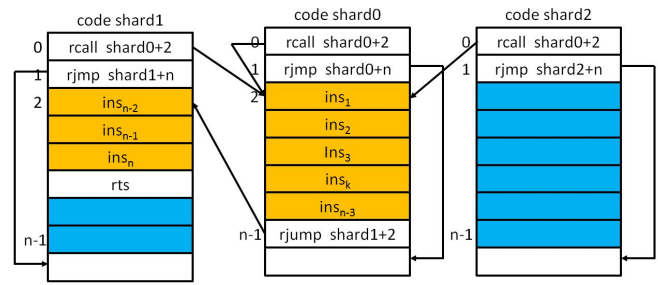

Fig5. Illustration of code compression, using duplicated code shards; freed memory space is colored in blue

As illustrated by Figure 5 (using instruction mnemonics from AVR processors), shards can be compressed by CALL, JUMP, and RTS instructions. The shard instructions are inserted in a subroutine, ended by RTS; the execution time is increased, but some extra memory is available for malicious code. An example of the result is presented in Figure 6.

```
610: 01 d0     rcall 0614       652: e0 df     rcall 0614
612: 1e c0     rjmp  0650       654: 1e c0     rjmp  0692
614: 3f 4f     sbci r19, 0xFF   656: 2c 91     ld   r18, X
616: d9 01     movw r26, r18    658: f2 26     eor  r15, r18
xyz:                            65a: 9f 01     movw r18, r30
64c: 17 96     adiw r26, 0x07   065c: 08 95    ret
64e: 03 C0     rjmp 0656        xyz:
650:                            692:
```

Fig6. Illustration of code compression, using duplicated code shards; which frees 52 bytes

### B. Threats

As mentioned above, hot wallets require online technology. They are therefore exposed to all the security threats cloud providers face, the most critical being secure key storage and their remote use. It should be noted that insider attacks are predominant and require a specific physical and logical security policy.

Cold wallets are obviously not exposed to such attacks; however, they are subject to threats that are listed in the following (non-exhaustive) list:

- T1) Lack of tamper-resistant storage and countermeasures for side-channel attacks, which allows cryptographic keys to be recovered at runtime;
- T2) Supply chain attacks aimed at malware injection or malicious firmware modification;
- T3) Software integrity is not verified. A bootloader checks the integrity of the updated firmware. It can be corrupted, similar to rootkits;
- T4) PIN or password hacking. For example, a key logger captures the text entered on the keyboard;
- T5) Misuse of the device from a laptop or cell phone, running a Trojan horse or worm;
- T6) Cloning of genuine devices. Clones, with equivalent features, include hidden functions targeting the recovery of cryptographic keys.

### III. ABOUT THE CRYPTO TERMINAL

The crypto terminal [10] is a keystore for blockchain wallets, designed to prevent these threats, thanks to adapted countermeasures. Its main services are the generation of signatures or transactions, with a high level of security and trust. However, it requires an external software component running on a PC or mobile device, which collects the information necessary to generate the transaction under the user's control.

It is based on open software and hardware, namely Arduino and javacard 3.0.4, which means that many form factors are possible. The core of the security is a removable javacard [11], which generates, computes, and stores the account keys and performs the transaction signatures.

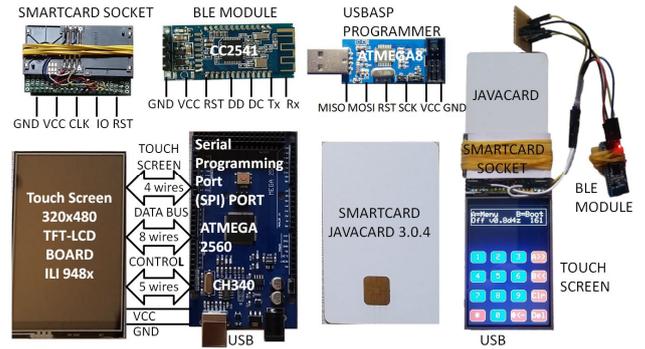

Fig7. Crypto terminal hardware components

### A. Open Hardware & Software

The crypto terminal (see Figure 7) comprises an 8-bit ATMEGA2560 microcontroller (16 MHz clock, 256KB FLASH, 8KB SRAM, 4KB EEPROM), a USB chip (CH340), a Bluetooth Low Energy (BLE) module [12] (CC2541, 256KB FLASH, 8KB SRAM), a 320x480 In-Plane Switching (IPS) touch screen (ILI 9486), a USBASP programming token [13] (ATMEGA8, 8KB FLASH, 1KB SRAM, 512B EEPROM), and a removable smart card (EAL6).

The BLE chip uses a five-wire interface for software download, the specifications of which are public. It can be flashed (up to 256KB) by open software, such as CCLOADER [24].

The USB programmer for Atmel AVR microcontrollers (USBASP) is an open project [13], including both hardware and software developments. It is supported by the open AVR Downloader/UploaDEr (AVRDUDE) initiative. Therefore, a complete platform is available for firmware download.

The ILI9486 is a device manufactured by *ILI Technology Corporation*. It includes a 40-pin interface with five control lines, an 8-bit data bus, and four wires dedicated to the resistive touch screen

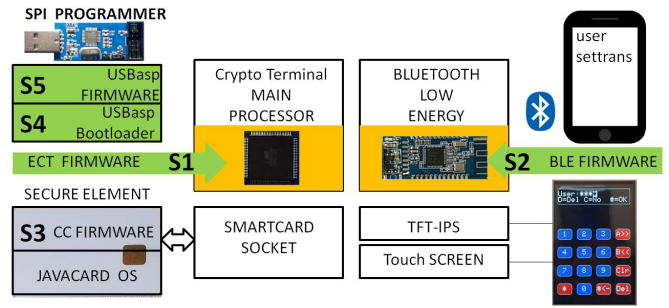

Fig8. Crypto terminal software components

The crypto terminal uses a set of five software packages (shown in Figure 8):
- S1) the main processor (AVR) firmware, up to 256KB;
- S2) the BLE module firmware, up to 256KB;
- S3) the firmware of the Java card;
- S4) USBASP bootloader;
- S5) the USBASP firmware, downloaded via the bootloader.

## B. Countermeasures against Cyber Attacks

We believe that trust in signatures is a key feature for the development of blockchain services. That is why we have developed and published technologies and algorithms to achieve integrity assurance for the software and hardware involved in signing transactions.

The security model includes the following elements:

C1) The security core is an EAL5+/ EAL6 removable security element (javacard), which stores the keys and performs the transaction signatures. Access to the smart card is protected by PINs.

C2) The authenticity of the smart card is guaranteed by an anti-cloning mechanism to detect "Evil Maid"-like attacks. The idea of such attacks is to use a cloned card to collect the user's keys. The public key of the smart card is signed by a certification authority (CA).

C3) The crypto terminal can duplicate the cryptographic contents of the secure element. The content of the smart card is self-attested (i.e., hashed and signed by the smart card).

C4) The crypto terminal acts as a firewall between the security element and the "hot" environment connected to the Internet. It also performs operations (key generation, signatures, etc.) in "cold" mode (i.e., without being connected to the Internet).

C5) Our security model is based on the "bare-metal" concept (i.e., the terminal is delivered without firmware and the Bluetooth module). Using a programming token, the user may flash these devices at any time.

C6) The firmware is authenticated by a built-in "Remote Attestation" algorithm. This algorithm [14] [15] [16] produces a result (which we call the authentication code) that cannot be predicted and requires a computation time depending on the result. A server can generate these codes, but so can the legitimate user, thus creating personal and unique authentication codes.

C7) Physical authentications of the cryptographic terminal processor and the programming token processor are performed using dynamic Physical Unclonable Function (PUF) techniques of SRAM (dPUF) [17]. A software memory probe extracts fingerprints from the SRAM after power-up. Innovative procedures have been developed to acquire such fingerprints, including singular points called flipping bits. Flipping bits [18] cannot be cloned by software. This technique ensures that devices received by clients are authentic.

C8) The programming token is a root of trust [19]. It manages the download of the processor firmware of the crypto terminal. We insert bootloader software [20] inside the programming token processor. The authenticity of this software is proved by a remote attestation algorithm generating authentication codes displayed by flashing LEDs. As mentioned earlier, the physical identity (dPUF) of this token allows clone detection.

## IV. JAVACARD SECURITY ELEMENTS

The javacard runs a crypto currency (CC) application written in the javacard programming language [11]. These smart cards are available from multiple manufacturers, with evaluation assurance levels (EAL) ranging from EAL5+ to EAL6+, according to the terminology of the Common Criteria standards. In a nutshell, the CC applet is a keystore, which provides secure key storage and ECDSA signing on a secp256k1 elliptic curve. The CC applet supports two-factor authentication: the knowledge of two PINs enables two working modes. The administrator mode gives access to all smart card services; the user mode allows reading public attributes and signatures.

According to Wikipedia, "*An evil maid attack is an attack on an unattended device, in which an attacker with physical access alters it in some undetectable way so that they can later access the device, or the data on it.*" A smart card software clone can include a backdoor. The principle of the "evil maid" attack consists in recovering the cryptographic keys of an unauthentic smart card.

To counter these threats, the CC application creates a private ($Priv_k$) and public ($Pub_k$) key pair during its instantiation. The public key is the identity of the device. The hash of this value is signed by a certification authority (CA), according to the ECDSA algorithm and the private key $Priv_{CA}$. So a tuple of two integer values (r,s) realizes the device certificate:

$$Cert_k = (r,s) = ECDSA_{Priv_{CA}}(SHA256(Pub_k))$$

In order to authenticate a genuine smart card, the crypto terminal performs the following procedure:
- 1) Reading the public key, $Pub_k$
- 2) Reading the certificate of the public key, $Cert_k$
- 3) Verifying $Cert_k$ with the CA's public key, $Pub_{CA}$
- 4) The knowledge of $Priv_k$ is proven through a challenge/response procedure. The terminal generates a random value (rnd) signed by the javacard private key:

$$ECDSA_{Priv_k}(rnd)$$

This signature is verified with $Pub_k$.

The self-attestation of the smartcard contents is produced according to the following procedure:
- 1) The crypto terminal generates a random value. The CC application calculates a hash value of its contents, concatenates this value with the random value, and generates a signature with its private key. The hash is returned with the signature; the response contains the following data:

$$hash \;||\; ECDSA_{Priv_k}(hash \;||\; rnd)$$

-2) The crypto terminal verifies the signature and, if successful, displays the hash (i.e., the fingerprint of the smart card's contents).

## V. TERMINAL SECURITY ELEMENTS

The firmware of the main processor controls the data exchange with the secure element. PIN entry from the touch screen prevents key logger attacks on the laptop or smartphone. All security-sensitive operations, such as signing, setting or exporting keys, are confirmed by the user to prevent unwanted actions by malware.

In the bare-metal approach, embedded firmware is erased and fully downloaded. Firmware flashing should prevent supply chain attacks, as it provides implicit proof of software integrity.

The programming sequence for the crypto terminal is as follows:

- 1) The main processor code is downloaded via an ICSP (In-Circuit Serial Programming) port. This firmware stores the certification authority's public key (used to authenticate the smart card) and a loader for the BLE module.
- 2) The BLE module is flashed using the USB interface and the built-in dedicated loader.

It should be noted that the touch screen device (ILI 9486) also includes a microcontroller unit (MCU), whose internal firmware cannot be updated. Nevertheless, the main processor only transmits data to this device and controls communication, thus reducing the attack surface.

The main processor firmware incorporates a remote attestation algorithm (detailed in Section 6) and is also capable of dumping and hashing the BLE module code.

The programming token (USBASP) stores a 2KB bootloader. Its integrity is proven by a remote attestation procedure detailed in Section 6.

## VI. REMOTE ATTESTATION

Integrity checking at runtime is a major security issue. Bare-metal functionality allows firmware to be downloaded at any time, so there is always a need to ensure the integrity of the software. To achieve this goal, we have designed a dedicated remote attestation algorithm called bijective MAC (bMAC).

Remote attestation is a process by which a trusted entity (the verifier) remotely measures the internal state of an untrusted and possibly compromised device (the verifier).

The bMAC verification is a self-verifying hash code (i.e., an instruction sequence that computes a fingerprint on itself and the memory contents), such that the MAC checksum is erroneous or the computation is slower if the instruction sequence or the memory contents are changed.

bMAC computes a fingerprint (h) on a set (A) of memories (FLASH, SRAM, EEPROM), whose size is m bytes, according to a pseudo-random order, fixed by a permutation P, such that:

$$bMAC(P) = h(\ A(P(0)\ ||\ ...A(P(i)\ ...\ ||\ A(P(m-1))\ )$$

The ICE algorithm presented in [14] computes a memory checksum according to a particular permutation. The permutation is an *Invertible Mapping* introduced in [25].

$$P(x) = x + x^2 \vee C \bmod 2^n$$

The least significant bit and the third bit of the constant C are both set to 1. In [14], n=16 (16-bit word) and C=5. The ICE algorithm has been implemented in a 16-bit von Neumann architecture with a hardware multiplier, leading to an optimized code size and execution time.

The paper [26] gives an exact characterization of the permutation polynomials modulo $n=2^w$, with $w \geq 2$

$$P(x) = a_0 + a_1 x + \cdots + a_d x^d \bmod 2^w$$

P(x) is a permutation polynomial if and only if $a_1$ is odd, the sum ($a_2+a_4+a_6+\ldots$) is even, and the sum ($a_3+a_5+a_7+\ldots$) is even.

The paper [27] demonstrates that:

$$P(x) = 1 + x + x^2 + \cdots + x^d \bmod(p^e)$$

is a polynomial permutation in field F(q), with $q=p^e$ and p prime, if and only if:

$$d = 1 \bmod(q(p-1))$$

We use the SHA256 or KECCAK-256 procedures for the MAC. The permutation P is based on exponential functions in the group Z/pZ*, with p a Saint Germain prime (p=2q+1, with q prime, p>m), and p=7mod8, which allows us to deterministically compute the generators ($g_k$)

$$g_k = p - (2^k \bmod p), k \in [1, q-1]$$

The P permutation is written as:

$$P(y) = F(1+y) - 1, y \in [0, p-2]$$

With:

$$F(x) = g_2^{s_1 g_1^x} \bmod p, x, s_1 \in [1, p-1]$$

A pseudo-random generator, detailed in [28], produces the parameters $g_2$, $g_1$, and $s_1$ from an integer value (31 bits) called SEED. Then, the bMAC function produces a MAC value.

The term $z=s_1 \cdot g_1^x$ is computed by a simple recursive procedure (i.e., $s_1 g_1^{x+1} = g_1 s_1 g_1^x \bmod p$). It is serialized as a bit stream ($b_i$) of n bits. The term $g_2^z$ is computed according to a *square and multiply* algorithm:

$$g_2^z = \prod_{i=0}^{n-1} g_2^{2^i} b_i \bmod p, \ b_i \in \{0,1\}$$

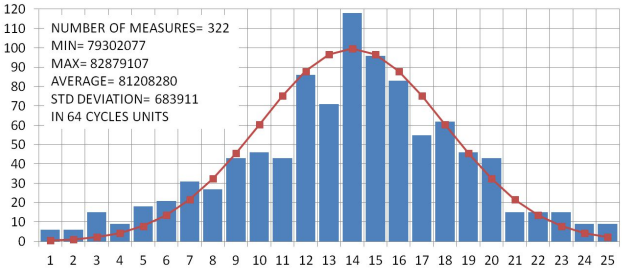

Fig.9 Distribution of BMAC computing time, for ATMEGA2560, 256KB, 16MHz clock

We observe that the bMAC computation time (cT) follows, roughly speaking, a normal distribution (see Figure 9) as a function of the SEED value. We believe this distribution is induced by the multiplication and modulus computations, depending on specific integer values.

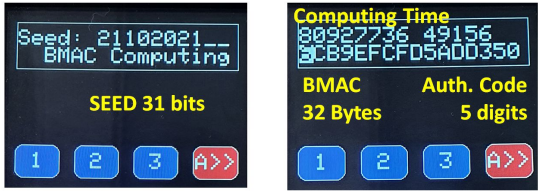

Fig.10 BMAC authentication code

The authentication code is a 16-bit integer value constructed with the two least significant bytes (see Figure 10) obtained from:

$$bMAC(SEED)\ exor\ cT$$

An internal timer measures the computation time. The idea of the *stop&start* attack is to stop this timer to perform hidden operations and then restart it. The timer uses a sub-clock (N=64) of the processor ($F_{clock}/N$), which protects against such an attack because the stop operation creates a random error in the range 0 to $T_{clock}$ (N-1).

We performed a *stop&start* attack by inserting instructions into the BMAC code that stop the internal clock and then restart it. This programming sequence has a prefix STS (short for Store Direct to data Space), which costs 2 cycles, and a suffix LDI+STS (LDI short for Load Immediate), which requires 3 cycles (1+2). For a memory size m, we should expect an increase in the measured time (i.e., number of cycles) of T=5.m. Experimental results with about 1000 samples show a normal distribution; on average, the increase in computing time is about T= m x 2.07 cycles (i.e., T/m = 2.07 cycles). The timer is stopped after the first STS instruction, which creates a (false) delay of 61, 62 or 63 cycles with a probability of about 1/64. Therefore, an estimate of T/m is:

$$\frac{T}{m} \# 5 * \frac{61}{64} - \frac{61}{64} - \frac{62}{64} - \frac{63}{64} = 1{,}86\ cycles$$

| Memory size | 8 KB | 32 KB | 256 KB |
|---|---|---|---|
| Average time in 64 cycles | 1.735.563 | 6.782.679 | 81.252.448 |
| Std deviation in 64 cycles σ | 778 | 3.122 | 696.043 |
| log$_2$ (σ) | 9,6 | 11,6 | 19,4 |

Fig.11 BMAC computing time average and standard deviation for three devices (ATMEGA8, ATMEGA368, ATMEGA2560) with different memory size.

We use AVR processors with a 16 MHz clock. Figure 11 shows the average computation time and standard deviation for three devices: ATMEGA8 (8KB), ATMEGA368 (32KB), and ATMEGA2560 (256 KB). The average is proportional to the memory size, about 1s/KB. The logarithm of the standard deviation seems to be proportional to the memory size; in other words, the BMAC entropy increases with the memory size.

VII. DYNAMIC PUF

The sale of hardware and software clones, which are not original devices, is a critical issue in complex supply chains. Since microcontrollers incorporate static RAM, we have developed authentication procedures based on the SRAM Physical Unclonable Function (SRAM-PUF). The SRAM-PUF is a physical identifier that can be used to authenticate the main processor of the crypto terminal (ATMEGA2560) and the programming token processor (ATMEGA8).

First, a memory probe firmware is downloaded into the processor, which also includes a UART interface. Second, the board, including this device, is powered via the ICSP port, with a controlled rise time. Third, the content of the SRAM is extracted via the serial interface.

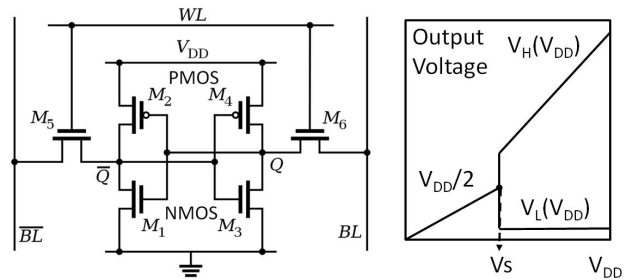

Fig.12 SRAM memory cell (left part), output voltage versus input voltage (right part).

On power-up, some SRAM cells take a fixed value. This effect is induced by the physical and electrical asymmetry of the PMOS and NMOS transistors (see Figure 12, left-hand side).

During power-up (see Figure 12, right-hand side), the output voltage of an SRAM cell remains close to VDD/2 until the gain is sufficient to switch to VL (logic low) or VH (logic high). For a highly mismatched cell, the output always takes a fixed value. The voltage (Vs) at which the transition occurs depends on the cell [17][18].

The paper [18] demonstrates the effect of voltage ramps on the PUF SRAM. Given a ramp V(t)= t VDD/T (i.e., slope=VDD/T), flipping bits are observed for slopes of high values. Flipping bits are created by capacitance mismatch, while most PUF-bits are due to voltage threshold differences (VTH). In other words, voltage ramps reproductively switch the value of some SRAM cells on power-up. A test RAM chip was designed with 180nm technology and simulated. Bit flipping was observed for T values below 15ms.

Thanks to the firmware memory probe, we read 1KB of SRAM. We perform these operations N times to extract the PUF cells. The static authentication of the processor requires only one reading of the SRAM, which is compared to a reference; the observed error is about 0.1%.

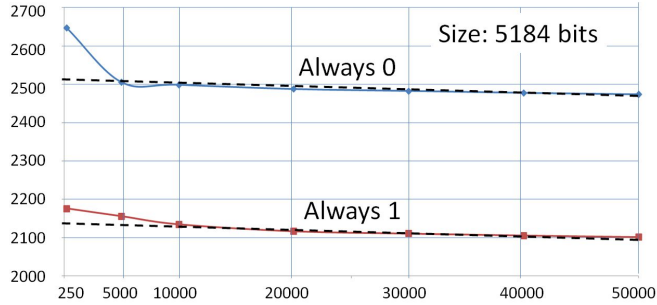

Fig13. SRAM PUF measurement results obtained with batches of 250 consecutive tries.

These values decrease slightly, about 5% for N between [250, 10000], and about 2% for N between [10000, 50000]. In our experiment, we record a set of 250 measurements in about 10 minutes; these data are aggregated to define N results (i.e., N/250 records). These results suggest that errors (i.e., an erroneous measurement of the state of a cell) occur randomly and increase with the duration of the measurement, like a Poisson distribution, according to a probability density function:

$$\rho(t) = \lambda\, e^{-\lambda t}$$

t being the measurement duration. What leads to an error probability:

$$p(t) = 1 - e^{-\lambda \cdot t} \# \lambda \cdot t$$

So the number of cells always seen at zero or one decreases like:

$$n(t) = n_0\,(1 - \lambda \cdot t)$$

Some PUF cells (about 5%) are sensitive to the voltage rise time and are called "flipping bits." For a "high" slope (>200mV/s), they have a content $b_k$, which is switched to ($1-b_k$) for a "low" slope (<10mV/s). We define [17] Sy waveforms constructed with two slopes (see Figure 14, right-hand side), switching at voltage y, y=0 corresponding to zero and y=1 to VDD. Therefore, S1 is the first slope, and S0 is the second. S0 (high slope) creates a flipping bit; S1 (low slope) does not create a flipping bit. We perform 25 measurements for each power-up with Sy, which allows us to estimate the threshold value y for each flipping bit. Figure 14 (left-hand side) shows the results for some flipping bits; we observe a low y threshold value.

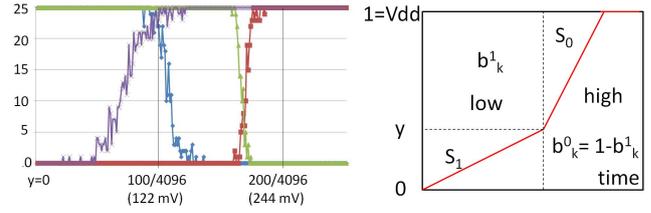

Fig.14 Flipping-bits state according to y value for 25 measures (left part), Sy powering-up signal (right part)

Dynamic processor authentication is based on flipping bits. We use the Slope & Square power-up waveforms [17] to set the state of the flipping bits. Rf is a square power-up waveform that creates flipping bits. We define the *Slope & Square* (Rs) waveform with a small slope of 625/512 mV/mS up to 512mS and then a fast rise time (see Figure 15, bottom), which does not create flipping bits. Since the processor does not operate at the threshold voltage of the flipping bits (<500mV), it cannot predict their value at runtime.

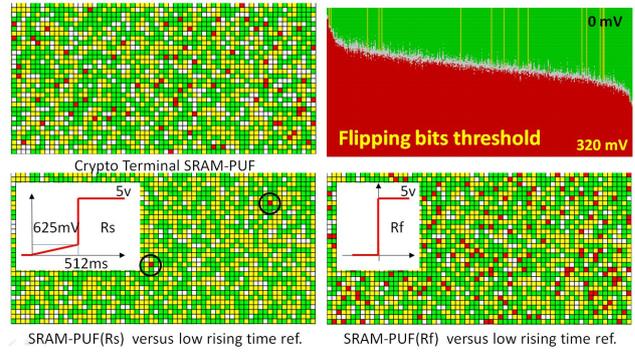

Fig15. Dynamic PUF (dPUF) (see text for comments)

Figure 15 illustrates the static and dynamic authentication for the main processor of the crypto terminal.

The upper left corner shows the result of 250 power-ups. SRAM cells always read to one are colored in green, those always read to zero are colored in yellow, and noisy bits (not always one or zero) are colored in white. The flipping bits observed for the Rf waveform versus the Rs waveform are colored in red. This bitmap is a kind of fingerprint of the processor.

The upper right corner shows the threshold voltage distribution for the flipping bits, which switch below the processor operating voltage at low voltage.

The lower left corner displays the comparison of the SRAM content for a single Rs waveform, with the reference obtained after 250 Rs. Only two errors (colored in red) are observed.

The lower right corner shows the comparison of the SRAM content for a single Rf waveform, with the reference obtained after 250 Rs. Many errors (colored in red) are observed.

## VIII. PROGRAMMING TOKEN AS ROOT OF TRUST

The USBASP programming token is built on an ATMEGA8 microcontroller, with 1KB SRAM, 2KB FLASH for the bootloader, and 6KB for the firmware. By default,

USBASP programmers do not have a bootloader. We use the bootloader for three reasons:

- To verify the bootloader's integrity using an integrity probe. This firmware runs the bMAC algorithm, which calculates the authentication code displayed by blinking LEDs.
- To authenticate the microcontroller, thanks to a memory probe and dPUF measurements. First, two references are collected with the power-up waveforms Rs and Rf. Then, memory dumps are performed with the Rs and Rf waveforms.
- To download the firmware required by the USBASP driver on the laptop side (e.g., Windows or Linux).

On the laptop side, the bootloader runs under the USBASP protocol. Therefore, software such as AVRDUDE can be used to download firmware to the MCU; the programming token is able to flash itself.

The bootloader is not activated by default, so the loaded firmware is executed. The bootloader is activated by a shortcut with the ground, with a time slot of five seconds, during which the internal flashing operation is activated.

It should be noted that programming tokens can program each other, so a trusted USBASP card can flash untrusted devices.

## IX. CONCLUSION

In this article, we described a crypto terminal based on open hardware and software technologies. It is equipped with a set of countermeasures to thwart cyber attacks against blockchain wallets. These innovative procedures, such as remote attestation or dynamic PUF, could also be deployed for other use cases, typically involving secure elements and an associated terminal.